\documentclass[journal]{IEEEtran}

\makeatletter

\newcommand{\Rmnum}[1]{\expandafter\@slowromancap\romannumeral #1@}
\makeatother

\usepackage[utf8]{inputenc}
\usepackage{graphicx}
\usepackage{subfigure}
\usepackage{cite}
\usepackage{multirow}
\usepackage{slashbox}
\usepackage{diagbox}
\usepackage{amssymb}
\usepackage{array}
\usepackage{slashbox}
\usepackage{gensymb}
\usepackage{bm}

\ifCLASSINFOpdf
\else
\fi
\begin{document}
%
\title{A Review on Serious Games for ADHD}
%
%
%

\author{Yuanyuan~Zheng, Rongyang~Li, Sha~Li, Yudong~Zhang, Shunkun~Yang,\\and~Huansheng~Ning,~\IEEEmembership{Senior Member,~IEEE}
}

\maketitle

\begin{abstract}

Attention deficit and hyperactivity disorder (ADHD) have two main characteristics: inattention and impulsivity.It has many obstacles to the normal development of children and is very common among children.As an intervention, Serious games for ADHD(SGADs) have shown great potential and are very effective for these ADHD patients.Although many serious games have been developed for ADHD patients, but a review paper that summarizes and generalizes the topic of video games has not yet appeared.In this article, we first classified serious games about ADHD according to different platforms developed by video games, and then we conducted a systematic review of video games that can help children with ADHD diagnose and treat. Finally, we discussed and made suggestions based on the current development of SGADs.
\end{abstract}


%
\IEEEpeerreviewmaketitle

\section{Introduction}
%
%
%
%
\IEEEPARstart{A}{ttention} deficit and hyperactivity disorder (ADHD), also known as hyperactivity, is mainly manifested as symptoms of inattention, impulsivity and hyperactivity.Although the prevalence of ADHD varies slightly according to different diagnostic criteria, it is generally about 7.2\% \cite{RE-1}.And boys have a higher prevalence rate than girls \cite{RE-2}.It can be seen that many children suffer from ADHD.
\footnote{Yuanyuan Zheng, RongYang Li, Sha Li, Huansheng Ning are with the School of
	 Computer and Communication Engineering, University of Science and Technology Beijing, 10083, Beijing, China, e-mail:ninghuansheng@ustb.edu.cn.
Huansheng Ning is also with Beijing Engineering Research Center for Cyberspace Data Analysis and Applications, 100083, Beijing, China}
\footnote{Yudong Zhang is with the School of Informatics, University of Leicester, UK, e-mail:yudongzhang@ieee.org.}
\footnote{Shunkun Yang is with the School of Information, University of Beihang, Beihang, Beihang, e-mail:ysk@buaa.edu.cn.}

Attention deficits cause children with ADHD to be unable to concentrate, have short attention spans, or be vulnerable to external interference. Hyperactivity disorder causes children with ADHD to have poor inhibition, difficult to control their emotions and behaviors, and easily impulsive.These symptoms of ADHD cause them to be at a disadvantage in many aspects of daily life.Children with ADHD have many academic problems, such as they can't complete their homework independently, can't concentrate on listening in class and are easy to drop out of school \cite{RE-3}.Moreover, children with ADHD often have language expression and reading comprehension barriers, which are very detrimental to their normal growth \cite{RE-4}\cite{RE-5}.Children with ADHD also exhibit some extreme behaviors in society, such as being prone to disputes with their relatives and friends, and being unable to get along with others normally.In addition, children with ADHD also bear the risk of other mental illnesses, such as conduct disorder, oppositional defiant disorder, depression, etc \cite{RE-6}.

Usually people use medication to relieve the symptoms of ADHD, but medication is prone to dependence and side effects \cite{RE-7}.With the development of science and technology, people began to use new technology to complete the intervention of ADHD. Therefore, serious games that achieve the purpose of training in the form of entertainment have been successfully applied to the treatment of ADHD.Nowadays, there are many serious games for ADHD, and these serious games have different effects due to different technologies.Through research, it is found that serious games about ADHD are roughly divided into two aspects: diagnosis and treatment, and they use different carriers to complete the successful operation of the game.In this paper, from the two aspects of diagnosis and treatment, we reviewed the SGADs, and analyzed and compared their technology and effect, hoping to let people know what impact serious games can bring to ADHD.

In the research of SGADs, according to the different carriers of video games, this paper divides SGADs into three categories: console games, computer games and mobile games.Console games can be played by connecting to a monitor such as a TV, and with the development of somatosensory interaction technology, players can perform a lot of physical movements when using gamepads or simulators to operate the game, which is not only easier to operate, but also to exercise.Console games can allow ADHD patients to devote themselves to the game, exercise while entertaining, and improve the coordination of the whole body.Computer games have more types, richer content and more realistic scenes. Most SGADs are developed for computers, and computer games developed for patients with ADHD can achieve a variety of therapeutic effects.The combination of computer games and brain computer interface (BCI)technology can detect the electroencephalogram(EEG) of ADHD patients, and the application of neural feedback technology can help ADHD patients with attention training.Serious games with BCI technology are different from general entertainment games, which involve complex brain wave knowledge.Therefore, Sung proposed a development framework for such games in his research, so as to reduce the difficulty of developing games \cite{RE-8}.Mobile games are popular with people for their portability and ease of operation. Most mobile games run on smart phones and tablet computers. People use their fingers to click and drag on the touch screen to complete the operation of the game.At present, there are many SGADs developed for mobile platforms so that patients with ADHD can be treated anytime and anywhere.

Serious games can play different roles in patients with ADHD. In this article, we analyze SGADs from the perspective of diagnosis and treatment.Usually, clinicians will use some questionnaires, statistical manuals or structured interviews to assess the daily behavior of patients, so as to determine whether they have ADHD \cite{RE-9}.However, this diagnosis method can easily be mixed with the subjective opinions of the evaluator and affect the reliability of the diagnosis. When children with ADHD are in this special environment of the hospital, their mental state is also easily affected.Faced with doctors and boring questionnaires, it is difficult for children with ADHD to have natural manifestations, which makes it more difficult to obtain an accurate diagnosis.Serious games can improve the user's participation in the evaluation process through immersive environment, interactive technology and multi-sensory experience to make the evaluation results more effective \cite{RE-10}.Gamified evaluation is more attractive than traditional evaluation, so that participants can complete the evaluation process without expending more energy, and it is possible to reduce the dropout rate during the evaluation process \cite{RE-11}.When using serious games to treat patients with ADHD, serious games can not only improve the attention and suppress impulse of patients with ADHD, but also exercise daily life skills and social skills of ADHD patients.Using serious games as an auxiliary tool can not only alleviate the symptoms of ADHD patients, but also improve the executive function of ADHD patients, and conduct cognitive training for ADHD patients \cite{RE-12}.All in all, for ADHD patients, serious games can play the role of diagnosis and treatment, and bring great hope to ADHD patients.

This article analyzes the effects that serious games can achieve in the diagnosis and treatment of ADHD patients. Our work has laid a good foundation for the development of serious games in ADHD in the future. The structure of the rest of this article is arranged as follows: Section \Rmnum{2} introduces SGADs developed for different platforms.section \Rmnum{3} gives a detailed overview of SGADs related to diagnosis and treatment. Section \Rmnum{4} discusses the advantages and disadvantages of these SGADs and gives suggestions. Section \Rmnum{5} summarizes the whole article.

\section{SGADs for different gaming platforms}
Usually, people call games that are not for entertainment purposes as serious games.Because serious games can achieve good training effects, serious games have achieved great success in many areas such as military, education, and medical treatment \cite{RE-13}.Because video games can provide constant stimulation and timely feedback, children with ADHD are often able to maintain their attention for a longer time when playing video games \cite{RE-14}.Therefore, the use of serious game therapy will allow ADHD patients to actively participate in the treatment process, thereby completing the training process smoothly and effectively.Video games can provide an immersive game environment, provide users with fascinating game challenges, and greatly stimulate the enthusiasm of players \cite{RE-15}.In order to allow researchers to better understand the development of SGAD, we divide SGADs into three stages: console games, computer games, and mobile games according to different game platforms (see Fig.1).And serious games at different stages also adopt different human-computer interaction technologies to complete the exchange of information with machines. These serious games can bring different gaming experiences to children with ADHD and help them achieve different training effects.

\begin{figure*}[t!]
	\centering
	\includegraphics[width=14cm,height=8cm]{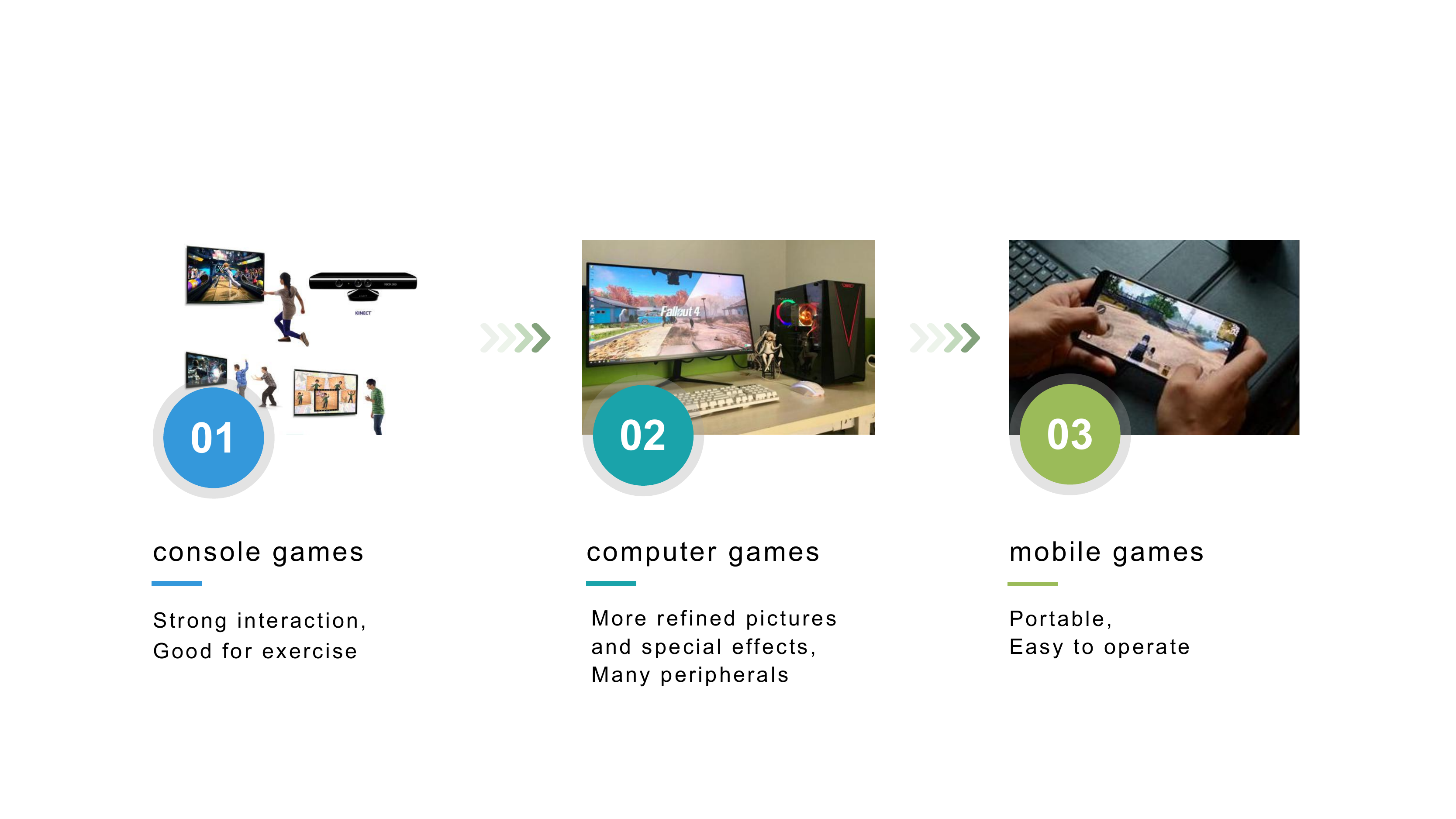}\\
	\caption{The development of SGADs }
	\label{Figure-1}
\end{figure*}

\subsection{console games}
Console games usually refer to the games on TV with the TV screen as the display. In order to play the corresponding console games, it needs to be equipped with game console, such as Nintendo's "switch" and Microsoft's "Xbox one".

Console games can use the interaction technology based on action recognition to complete the interaction with the game.Action recognition uses optical devices and acceleration sensors to sense the user's position and speed, determine the user's action, and then carry out interactive activities.This technology allows the console game to break through the previous operation mode of simple input by the handle button, but to operate through changes in body movements.Action recognition technology makes human-computer interaction in a more natural and intuitive way, which makes it more convenient for users to operate the game.

Console games can provide a highly interactive multiplayer cooperation mode, which can promote communication with friends and narrow the distance between family members.The game console with somatosensory function can also make people exercise while playing, not only can enjoy the rich game screen, but also can do a lot of physical movements, and the purpose of exercise can be achieved in the process of entertainment.

Wii is a game console released by Nintendo in 2006. Its game sensors enable players to control the characters in the game through actions.When children with ADHD perform operations, it can bring them a variety of sensory stimulation.Chuang conducted an experiment to study the effects of Wii play and Wii sport on ADHD children, and the results showed that ADHD children can feel more fun and improve their attention when playing games \cite{RE-16}.Children with ADHD often have defects in executive function, that is, it is difficult to complete tasks that require planning and organization \cite{RE-17}.Use SGADs can train for specific executive function deficits, experiments show that SGADs can increase the enthusiasm of children with ADHD to participate in treatment and improve the effect of executive function training \cite{RE-18}.

As an auxiliary measure, physical exercise can be used to reduce the executive function defects of ADHD children. Because exercise requires a lot of energy and energy, ADHD children are often reluctant to take physical exercise.Console games that can be used for sports activities gamify sports and cognitive training, making sports more interesting and challenging, and can exercise players' concentration, inhibition, and reaction. Benzing \cite{RE-19} used Microsoft's Xbox game console and equipped with Kinect equipment capable of motion recognition to conduct an intervention experiment on selected patients. During the experiment, the experimental group used Shape UP and Beatmaster Training Quest for training. A questionnaire was used to evaluate the symptoms of ADHD in children before and after the intervention. The evaluation results show that exercise games can not only reduce the inhibition of these patients and improve their responsiveness, but also improve the deficits in executive function.

\subsection{computer games}
Computer games refer to electronic games played on computers. Video games are divided into client games and web games.Client games need to download the client, the game screen and special effects are more refined and the higher the client computer configuration, the more refined the game scene.Web games can be entered from a computer browser, without downloading the client, you can play with a computer and the Internet, which is more convenient than client games.Computer games can be connected to multiple devices to achieve different game effects.

Computer games generally use the mouse and keyboard to complete the interaction with the game.Mouse and keyboard are the most common hardware input devices in people's daily life. Most electronic games rely on the mouse and keyboard to complete the transmission of information in the process of human-computer interaction.This control method occupies the mainstream position in computer games. As a traditional control method, although it can meet the general needs of the public, it cannot bring users a natural interactive experience.Such video games can help children with ADHD improve their attention, suppress impulse, and train daily life skills.

Children with ADHD often have symptoms of inattention. Studies have found that the level of attention is related to brain waves. So people began to apply brain-computer interaction technology to SGADs based on computer games.BCI can be used to complete the communication between the brain and the computer. In BCI technology, EEG equipment collects brain electrical signals and directly inputs brain activity to the computer, so no input devices such as a mouse and keyboard are needed to complete communication \cite{RE-20}.The two most popular commercial brain-computer interface devices on the market, Emotiv and NeuroSky, have been successfully applied to the control of electronic games \cite{RE-21}.

According to different frequencies, EEG signals are divided into different frequency bands, which also correspond to different brain activities.The ratio of $\theta$/$\beta$ in the EEG is related to attention. The higher the ratio, the less concentrated the attention, and the higher $\theta$/$\beta$ ratio  has become a biological characteristic of patients with ADHD \cite{RE-22}.Therefore, in order to make the attention of ADHD patients more concentrated, training can be used in video games to reduce the ratio of brain waves $\theta$/$\beta$ \cite{RE-22}.In video games, the $\theta$/$\beta$ ratio can be improved by applying auditory and visual stimuli to patients with ADHD. Studies have shown that light music has the best effect and can effectively reduce the $\theta$/$\beta$ ratio \cite{RE-23}.

Neurofeedback uses BCI equipment to collect the user’s EEG signals when playing video games, and immediately feedback them to the user, helping children with ADHD train their brainwave activities \cite{RE-24}.Someone conducted a randomized controlled trial on children with ADHD, adopted different intervention measures for different experimental groups, and used questionnaires and scales to evaluate the intervention of ADHD children to verify the effectiveness of neurofeedback in the treatment of ADHD \cite{RE-25}.Evidence shows that neurofeedback can effectively improve the symptoms of ADHD \cite{RE-26}.The factors that affect the efficacy of neurofeedback have also been fully studied. Experiments have shown that the depth of treatment, evaluation reports provided by different observers, and different EEG equipment can all affect the effect of neurofeedback therapy \cite{RE-27}.

In order to improve the player's visual experience, the combination of computer games and virtual reality technology brings players an extraordinary gaming experience \cite{RE-28}.Virtual reality technology changes the traditional way of operating games with a mouse and keyboard, and is committed to improving the authenticity and interactivity of games.Serious games using virtual reality technology can provide an immersive environment and help transfer the skills learned during training to real life. The serious game of virtual reality provides a controllable reality environment for cognitive exercises and can provide timely feedback, which is conducive to improving the symptoms of ADHD \cite{RE-29}.Alqithami \cite{RE-30} proposed a virtual reality therapist system that combines virtual reality and agent-based models.The biggest advantage of virtual reality technology is that it can exercise the players' reaction ability and sensitivity in the game process, so that they can quickly integrate into the training environment.

Although virtual reality technology can bring players a more realistic interactive experience, computer games based on this technology also have disadvantages such as high cost, single game category, and easy physical discomfort. Moreover, the head-mounted devices used in such computer games are not suitable for children, so the application of computer games based on virtual reality technology in ADHD has yet to be developed.
\subsection{mobile games}
Mobile games use touch-based interactive technology to complete information input.This technology is often used on mobile devices such as mobile phones and tablets that require touch control.Mobile games are easier to operate and adapt to a wider range of people, so mobile games are more humanized than computer games.Users can use their fingers to complete operations such as clicking, dragging, and so on, which makes the operations of some electronic games easier and enhances the user's gaming experience.

Compared with computer games, portability is the most prominent feature of mobile games. You can play games anytime, anywhere with only a mobile device.Generally, mobile games are easy to operate and suitable for a wide range of people, but the animation effect of mobile games is not as strong as computer games, and some special effects are not as good as computer games.Mobile games can also be played on a tablet. Compared with a mobile phone, a tablet can provide a more open visual effect.The biggest advantage of a tablet is a larger screen, higher resolution, and better picture quality.Tablet computers can make it easier for players to participate in games, and are more eye-protective than mobile phones, but due to their larger size, they are not easy to operate for some games.

\begin{figure*}[t!]
	\centering
	\includegraphics[width=14.63cm,height=8.23cm]{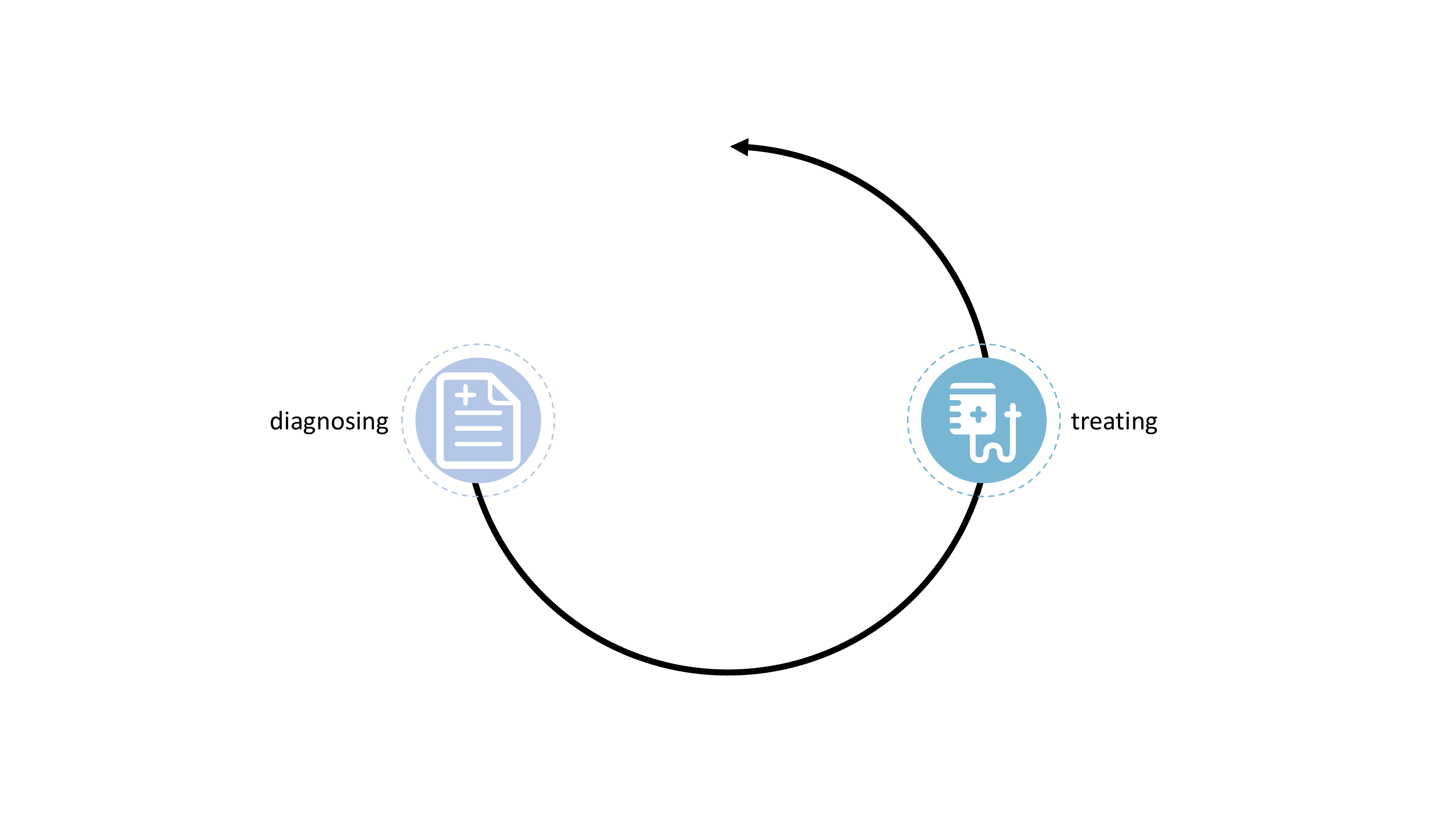}\\
	\caption{The role of SGADs }
	\label{Fig-2}
\end{figure*}

\begin{table*}\normalsize
\newcommand{\tabincell}[2]{\begin{tabular}{@{}#1@{}}#2\end{tabular}}
\renewcommand\arraystretch{2}
	\centering
	\caption{Advantages and disadvantages of SGADs}
	\label{Tab-1}
	\begin{tabular}{|m{2cm}<{\centering}|m{6cm}<{\centering}|m{8cm}<{\centering}|}\hline
		{Types}&{Diagnosing}&{Treating}\\\hline
		\tabincell{c}{Advantages}&\tabincell{c}{High participation rate,\\High accuracy rate,Objective}&\tabincell{c}{Low side effects,Increase enthusiasm,\\Improve executive function,\\Improve daily life skills,Promote social interaction}\\\hline
		\tabincell{c}{Disadvantages}&\tabincell{c}{Cannot distinguish subtypes clearly}&\tabincell{c}{The effect is not obvious,\\Easily addictive,Harmful to eyesight}\\\hline

	\end{tabular}
\end{table*}

\section{SGADs with different purposes}
Table1 summarizes the advantages and disadvantages of SGADs.

As an auxiliary tool, serious games play the process of diagnosis and treatment of ADHD children (see Fig.2).A large number of experiments have proved that using video games can effectively distinguish ADHD children from non ADHD children, and it also has a very obvious effect in reducing the symptoms of ADHD children and improving executive function \cite{RE-31}.Although serious games can bring a lot of positive effects to ADHD children, there are also some limitations.

\subsection{SGADs for diagnosing ADHD}
The research projects of serious games related to diagnosis are shown in Table 2.

Usually, medical staff complete the diagnosis of ADHD children by analyzing the results of a questionnaire about the psychological test for ADHD filled by teachers and parents.This method is not only boring and complicated for children with ADHD, but also because the questionnaire is filled out by parents and teachers, it is also prone to inconsistent results of the questionnaire, which is not conducive to accurate and objective diagnosis.Therefore, the diagnosis process is gamified, and the diagnosis result is completely dependent on the performance of ADHD children in the game. This can not only stimulate the fun of ADHD children to participate in the diagnosis, but also make the diagnosis results more objective.

Khaleghi \cite{RE-32} has developed a computer game for diagnosing and evaluating ADHD, which uses DSM-IV as the diagnostic criteria.The diagnosis result depends on the number of times the player uses the mouse to click on irrelevant answers in the game, the answers chosen, and the time spent.There is also a computer game used to diagnose ADHD, its name is Supermarket \cite{RE-33}.The biggest feature of this video game is that the data obtained by children with ADHD during the game is classified by data mining algorithms. By using machine learning technology to classify data, it can be used as an auxiliary means to correctly distinguish ADHD children from non-ADHD children.Chen \cite{RE-34} designed a diagnostic aid system COSA, this system evaluates the data obtained in the game according to the DSM-V diagnostic criteria.This system can not only test physiological data, but also evaluate the movement data generated by the player during the game, and the effectiveness of this system has also been verified.Because ADHD children tend to have a lower enthusiasm when facing traditional cognitive tests, this is likely to affect the test results, so through gamification, children with ADHD can show the most natural state.Another computer game called Timo’s Adventure can be used to evaluate the deficits in executive function \cite{RE-35}.The system consists of six small games with different functions, and each small game can be used to improve different executive function defects. It is judged whether it is ADHD by the defects of executive function displayed during the game, and the effectiveness of this game is proved through experiments.

With the development of mobile devices, video games developed for mobile devices have gradually taken a place in the electronic game market.This type of video games is not limited to time and location restrictions. You can play video games anytime, anywhere with only a mobile phone or tablet.And players use their fingers to use the touch screen as the medium to complete the interaction with the device, which makes it easier to operate the game, so SGADs based on mobile devices came into being.Nayra \cite{RE-36} has designed a mobile application that combines the functions of pre-diagnosis and treatment.People related to ADHD can communicate through this application, and ensuring the data security of their communication is also a key part of the application.This application is also based on the questionnaire to complete the assessment of ADHD. The difference is that the results of the questionnaire are first classified by the naive Bayes classifier and then sent to the medical staff. The medical staff then complete the diagnosis of the child with ADHD based on the data.

Since inattention is the main feature of ADHD, it is possible to determine whether they have ADHD by testing the attention level of the child.Attention level can be given by EEG signals, and EEG signals can be obtained through brain-computer interface devices.This diagnosis method combines brain-computer interaction technology with serious games to obtain the EEG signal of the player during the game, and compare it with the characteristics of the EEG signal diagnosed with ADHD, thereby completing the diagnosis of ADHD .A serious game called FOCUS relies on brain-computer interaction technology to diagnose ADHD \cite{RE-37}\cite{RE-38}\cite{RE-39}.This video game not only uses brain-computer interaction technology to obtain EEG data, but also uses virtual reality technology to create an immersive game environment, and uses machine learning technology to classify the obtained data to accurately diagnose patients with ADHD .In the process of playing the game, the player's attention level will be promptly fed back to the player, and the player can adjust his attention in time according to the neurofeedback, so this game can also achieve the purpose of training attention.

A serious game that uses Kinect to recognize actions also completes the diagnosis of ADHD by measuring attention \cite{RE-40}.This is a 3D video game in which the player controls the game by moving his body, and the level of attention is given by the measured variables that the player obtains during the game.

Serious games can also use virtual reality and machine learning techniques to evaluate patients with ADHD.. On the one hand, virtual reality provides an immersive environment that can provide a real sensory experience during the diagnosis process. On the other hand, machine learning can classify the data obtained during the game and increase the accuracy of diagnosis. For example, combining virtual reality and deep learning, processing the obtained data through CNN(convolutional neural network), and according to the results of data processing to determine whether suffering from ADHD \cite{RE-41}.

Empowerment brain is a video game used to evaluate the symptoms of ADHD in autistic children \cite{RE-42}. This game uses augmented reality technology to make the game picture more realistic, and uses Google glasses to analyze whether the performance of autistic children in the game is related to ADHD by using artificial intelligence.

\begin{table*}\normalsize
\newcommand{\tabincell}[2]{\begin{tabular}{@{}#1@{}}#2\end{tabular}}
\renewcommand\arraystretch{2}
	\centering
	\caption{Characteristics of the reviewed studies: diagnosis}
	\label{Tab-2}
	\begin{tabular}{|m{3cm}<{\centering}|m{6cm}<{\centering}|m{5cm}<{\centering}|m{2cm}<{\centering}|}\hline
		{Game Name}&{Purpose}&{Technologies}&{References}\\\hline
		\tabincell{c}{AGE 1}&\tabincell{c}{Diagnosing cognitive impairment}&\tabincell{l}{PC}&\tabincell{l}{[32]}\\\hline
		\tabincell{c}{Supermarket}&\tabincell{c}{Distinguish ADHD from non-ADHD}&\tabincell{l}{PC}&\tabincell{l}{[33]}\\\hline
		\tabincell{c}{COSA}&\tabincell{c}{Assess inattention and impulsivity}&\tabincell{l}{Sensor, Display}&\tabincell{l}{[34]}\\\hline
		\tabincell{c}{Timo’s Adventure}&\tabincell{c}{Assess specific cognitive functions}&\tabincell{l}{PC}&\tabincell{l}{[35]}\\\hline
		\tabincell{c}{mHealth}&\tabincell{c}{Pre-diagnosis,\\Improve cognitive ability}&\tabincell{l}{Mobile,Tablet PC,Touch interaction}&\tabincell{l}{[36]}\\\hline
		\tabincell{c}{FOCUS}&\tabincell{c}{Detect attention level \\and improve attention}&\tabincell{l}{PC,BCI}&\tabincell{l}{[37],[38],[39]}\\\hline
		\tabincell{c}{-}&\tabincell{c}{Detect attention level \\and improve attention}&\tabincell{l}{Console,Somatosensory interaction}&\tabincell{l}{[40]}\\\hline
		\tabincell{c}{Empowered Brain}&\tabincell{c}{Detect symptoms of ADHD}&\tabincell{l}{PC,AR}&\tabincell{l}{[42]}\\\hline

	\end{tabular}
\end{table*}

\subsection{SGADs for treating ADHD}
The research projects of serious games related to treatment are shown in Table 3.

\begin{table*}\normalsize
\newcommand{\tabincell}[2]{\begin{tabular}{@{}#1@{}}#2\end{tabular}}
\renewcommand\arraystretch{2}
	\centering
	\caption{Characteristics of the reviewed studies: diagnosis}
	\label{Tab-3}
	\begin{tabular}{|m{4cm}<{\centering}|m{6cm}<{\centering}|m{5cm}<{\centering}|m{2cm}<{\centering}|}\hline
		{Game Name}&{Purpose}&{Technologies}&{References}\\\hline
		\tabincell{c}{Brain Chi,Pipe,\\Dancing Robot,\\Escape}&\tabincell{c}{Improve attention}&\tabincell{l}{PC,BCI}&\tabincell{l}{[43],[44],[45]
}\\\hline
		\tabincell{c}{Armis}&\tabincell{c}{Improve attention}&\tabincell{l}{PC,BCI}&\tabincell{l}{[46]}\\\hline
		\tabincell{c}{A 3D Learning Playground}&\tabincell{c}{Improve attention}&\tabincell{l}{PC,BCI}&\tabincell{l}{[47]}\\\hline
		\tabincell{c}{Harvest Challenge}&\tabincell{c}{Improve attention,Suppress impulse}&\tabincell{l}{PC,BCI}&\tabincell{l}{[48],[49]}\\\hline
		\tabincell{c}{Neurofeedback Space}&\tabincell{c}{Improve attention}&\tabincell{l}{PC,BCI}&\tabincell{l}{[50]}\\\hline
		\tabincell{c}{Keep Attention}&\tabincell{c}{Improve attention and memory}&\tabincell{l}{PC}&\tabincell{l}{[51]}\\\hline
		\tabincell{c}{Boogies Academy,Cuibrain}&\tabincell{c}{Improve attention}&\tabincell{l}{Mobile,Tablet PC,Touch interaction}&\tabincell{l}{[52]}\\\hline
		\tabincell{c}{ATHYNOS}&\tabincell{c}{Improve attention,\\time management and social skills}&\tabincell{l}{AR,Somatosensory interaction}&\tabincell{l}{[53]}\\\hline
		\tabincell{c}{Plan-It Commander}&\tabincell{c}{Improve time management,\\planning organization \\and social skills}&\tabincell{l}{PC,3D}&\tabincell{l}{[54],[55]}\\\hline
		\tabincell{c}{-}&\tabincell{c}{Improve time management \\and task prioritization skills}&\tabincell{l}{Mobile,Tablet PC,Touch interaction}&\tabincell{l}{[56]}\\\hline
		\tabincell{c}{Braingame Brian}&\tabincell{c}{Improve attention,Suppress impulse,\\Train executive function}&\tabincell{l}{PC}&\tabincell{l}{[58]}\\\hline
		\tabincell{c}{Antonyms}&\tabincell{c}{Suppress impulse}&\tabincell{l}{PC}&\tabincell{l}{[59],[60]}\\\hline
		\tabincell{c}{The Secret Trail of Moon}&\tabincell{c}{Improve attention,Suppress impulse,\\Train executive function}&\tabincell{l}{VR}&\tabincell{l}{[61]}\\\hline
		\tabincell{c}{Reading and Comprehension}&\tabincell{c}{Improve reading comprehension}&\tabincell{l}{Mobile,Tablet PC,Touch interaction}&\tabincell{l}{[62]}\\\hline
		\tabincell{c}{ADDventurous \\Rhythmical Planet}&\tabincell{c}{Improve social skills}&\tabincell{l}{Sensor equipment}&\tabincell{l}{[63],[64]}\\\hline
		\tabincell{c}{EmoGalaxy}&\tabincell{c}{Promote emotional regulation \\and improve social skills}&\tabincell{l}{PC,Mobile,Tablet PC}&\tabincell{l}{[65]}\\\hline
	
	\end{tabular}
\end{table*}

The main characteristic of ADHD is inattention, main performance in difficult to concentrate, and the duration of the attention is shorter, also very vulnerable to outside interference.This feature has a lot of negative effects on ADHD children's learning and life, such as they can't insist on completing tasks in life, it's easy to give up halfway in doing things and so on, they can't pay attention in class and listen carefully in learning, and they can't complete the homework assigned by teachers and so on.At present, although there are many measures can be used to improve the symptoms of ADHD, the effect is not obvious, and it is easy to produce side effects, such as drug intervention, which is not only ineffective, but also harmful to the health of ADHD children.However, serious game intervention can make up for these deficiencies. The combination of serious game and different human-computer interaction technology can bring a variety of interactive experience to patients with ADHD and achieve different effects in the treatment of ADHD.

Serious games based on brain computer interface have been gradually applied to the treatment of ADHD, whose main purpose is to improve the attention level of ADHD.By using some EEG acquisition equipment to obtain the EEG data of patients with ADHD in the game process, and then provide real-time feedback to patients through the screen or sound in the game, it can help patients with ADHD to train their ability of concentration \cite{RE-43}.In serious games based on EEG, EEG signal processing is usually implemented by some algorithms, such as frequency analysis and event-related potential analysis \cite{RE-44}.In order to accurately capture the different states of brain activity, some studies proposed EEG signal processing method based on fractal dimension model, and successfully applied to serious games \cite{RE-45}.

In a video game called armis \cite{RE-46}, players use keyboard to control the characters in the game, while wearing wireless EEG device to detect the state of the brain, and use the change of game background color to represent the level of attention in the process of playing the game. Through this intuitive visual experience, players can understand their level of attention more clearly and adjust their attention according to the requirements of the game.A BCI video game with a 3D virtual classroom as the background can also be used to train the player's attention \cite{RE-47}.This video game can effectively transfer the training effect to the real world by simulating the situation of students in the classroom.

A video game called harvest challenge also gets players’EEG through BCI device, and then trains players' attention through neural feedback \cite{RE-48}. This video game is composed of three small games, which can train ADHD children's ability of waiting, planning, following instructions and achieving goals \cite{RE-49}. The enhancement of these four abilities can help ADHD children to improve their attention and restrain their impulses.Neural feedback space is also a video game using BCI technology, and this video game takes the spaceship as the background, the head mounted device collecting EEG signals corresponds to the space helmet in the game, and the speed of the spaceship as the feedback signal represents the level of players'attention \cite{RE-50}.

ADHD patients can also play some other types of video games to exercise their attention.For example, Hocine has developed an electronic game controlled by a mouse and keyboard, which can also be used to exercise players’concentration and memory, and use feedback test questionnaires to help players adjust themselves during training \cite{RE-51}.Redondo conducted an experiment to study the effects of the two video games "Boogies Academy" and "Cuibrain" on visual attention \cite{RE-52}.These two electronic games are based on multiple intelligences and are suitable for mobile phones and tablets respectively. They can be used to activate language intelligence and interpersonal intelligence.ATHYNOS is a serious game that uses augmented reality technology \cite{RE-53}.It can not only improve attention, but also fully mobilize the hand-eye coordination of children with ADHD, reduce reaction time, and promote children with ADHD to participate in the treatment voluntarily.

The symptoms of inattention and impulsivity in ADHD children bring a lot of burden to their daily life.Plan it commander, as a typical serious game developed to exercise life skills, is very helpful for ADHD patients to carry out rehabilitation training \cite{RE-54}. Three small games are designed to improve the time management, planning organization and social skills of ADHD children. And a randomized controlled experiment was carried out to verify its effectiveness \cite{RE-55}.In the experiment, it was found that with the increase of training times, ADHD children's daily life skills were gradually improved.Pascual has developed a video game with a virtual balance as the background of the game \cite{RE-56}.This video game can drag and drop two different tasks to the two ends of the virtual balance on a touch-based interactive device, and then help users to prioritize the two tasks according to the decision tree.

Children with ADHD also have deficits in executive function \cite{RE-57}. Executive function belongs to a kind of cognitive function, which is a human inherent ability, including planning and organization, time management, working memory, reaction inhibition, etc.Braingame Brian is a video game mainly used for executive function training, and it can also be used to suppress the impulsivity of patients with ADHD  \cite{RE-58}.And the reward mechanism and strategy designed in this game also promoted the enthusiasm of ADHD patients to participate in training.The main function of antonyms is to suppress impulse, through this video game, ADHD patients can gradually learn to control themselves \cite{RE-59}.The player completes the interaction with the game scene through the touch screen. If the player wants to win the game, he must learn to control himself, adjust his emotions and suppress impulse \cite{RE-60}.The secret track of the moon is a video game with chess as the main mechanism, which uses virtual reality technology to bring players a more real game experience \cite{RE-61}. This video game can provide cognitive training for ADHD patients to improve their cognitive function.

Wrońska has developed a serious game to improve the reading comprehension ability, which is related to the faculty of memory in executive functions \cite{RE-62}.ADHD patients are easily affected by inattention when reading or looking at pictures, causing them to be unable to understand the content of the reading.This video game was developed for ipad. It uses touch-sensitive interactive devices to complete working memory training for children with ADHD, which is not only helpful for their daily life, but also for improving their academic performance \cite{RE-62}.

Children with ADHD are often under a lot of stress and easily lose control of their emotions.Through breathing exercises for ADHD children, they can reduce their stress and learn to adjust their emotions. Chillfish mainly uses LEGO fish with respiratory sensors to complete the interaction with video games \cite{RE-63}. By using the respiratory controller to breathe slowly, ADHD children can gradually keep calm and relaxed.Music can eliminate tension and relieve stress, and regular exposure to music rhythm helps promote physical and mental health.ADDventurous Rhythmical Planet is an video game that uses music rhythm for training \cite{RE-63}.Players control the drums with sensors according to the music rhythm provided by the game.One of the characteristics of this game is that it can also choose the multiplayer mode, which requires the cooperation between players. This video game can not only improve the attention of children with ADHD and inhibit their impulses, but also promote the cooperation and communication between children with ADHD and others \cite{RE-64}.EmoGalaxy is a video game that focuses on emotional regulation to improve social skills \cite{RE-65}. It is mainly used to complete the training of the three aspects of recognizing emotions, expressing emotions and regulating emotions. Players need to express the correct emotions on the planet representing different emotions to proceed to the next level, and the recognition of emotions is achieved using facial recognition technology.

\section{Discussion and suggestions}
Serious games as a new technology can provide great help to patients with ADHD. Different from traditional methods, serious games gamify the process of diagnosis and treatment, which can not only bring fun to children with ADHD but also reduce symptoms.When making a diagnosis, it is possible to accurately determine whether they have ADHD through the performance of the players in the game and some data collected, And can also use machine learning technology to classify game data to detect ADHD patients.Different game mechanisms and operation methods can bring different training effects to children with ADHD. These games have played a role in helping children with ADHD strengthen their attention, improve executive functions, and enhance social communication skills.

Because electronic games can provide players with rich game scenes and timely rewards, they greatly arouse the interest of children with ADHD.Most importantly, video games can simulate real scenes, and with the development of augmented reality and virtual reality, children with ADHD can be placed in a more realistic game scene. This is conducive to diagnosis and treatment based on their behaviors and reactions in the game, and the training effect achieved in the game is also conducive to transfer to real life.However, SGADs also have some shortcomings, as detailed below.
\begin{itemize}	
	\item  Addiction to online games.Although it is designed to improve ADHD, video games are still carried out in the form of entertainment, and most of the people with ADHD are children. Children themselves are easily attracted by video games. Therefore, when using video games to treat ADHD children, we should not ignore the problem of game addiction.In response to this problem, we recommend that the visual effects of the game design must not be too eye-catching, otherwise the bright colors are likely to cause interference to the players. The rules and gameplay of the game should also focus on cultivating concentration and suppressing impulse. The characters and scenes of the game should best correspond to real life, so that video games can play a real role.	
	\item  Different types of serious games were applied to different ADHD patients.The majority of ADHD patients are pre puberty children, and boys are the majority.According to the characteristics of the patients, the operation of SGADs should not be too complicated.Different types of games can be implemented according to different patients, such as dress up, leisure and board games for girls, shooting, sports and adventure games for boys.
	\item  Adapt commercial games into SGADs.In the development of serious games for ADHD, because they are not suitable for the majority of people, the cost is high and the development is difficult, so we can consider adapting some commercial games into sgads, which not only saves the cost, but also applies to a wider range of people.
	\item  Accurate classification of symptoms of ADHD.In some serious games for the diagnosis of ADHD, most of the games can only diagnose one symptom of inattention, and there is no strict classification of ADHD. It is hoped that researchers can design serious games to divide ADHD patients into three subtypes: inattention, hyperactivity and mixed type when developing serious games for the diagnosis of ADHD in the future.
	\item  Adjust the evaluation criteria.When serious games are used to diagnose ADHD, some questionnaires and evaluation scales are still used to evaluate the behavior of patients in real life. There is no research to confirm whether there are differences in the use of these standards to evaluate the performance of patients in games. Whether the evaluation criteria should be adjusted according to the diagnosis of patients using serious games remains to be studied.
	
\end{itemize}

\section{Conclusion}
Through the research and investigation of a large number of literatures, this article found that serious games have been able to achieve very significant effects in the application of ADHD.Unlike traditional intervention methods, the use of gamification has greatly improved the participation and enthusiasm of children with ADHD.This article starts from the different platforms for developing electronic games, studies the development process of electronic games, and makes a detailed analysis of the technology and significance adopted by them.In this article, we have conducted a systematic review of SGADs from the perspective of diagnosis and treatment, and detailed analysis of the game mechanism and operation methods of each game, as well as an overview of the specific effects that these games can achieve.Finally, we made a few suggestions, hoping to help SGADs develop better.In summary, our work can help people have a more comprehensive understanding of SGADs, and help researchers develop more effective SGADs.



\section*{Acknowledgment}
The work was supported by teaching reform of University of Science and Technology Beijing under JG2020Z08.
\bibliographystyle{unsrt}
\bibliography{Reference}

%

\end{document}